# Advancing computerized cognitive training for early Alzheimer's disease in a pandemic and post-pandemic world

Kaylee Bodner, Terry E. Goldberg, D. P. Devanand, P. Murali Doraiswamy

Neurocognitive Disorders Program, Departments of Psychiatry and Medicine,
Duke University School of Medicine
Division of Geriatric Psychiatry, New York State Psychiatric Institute and Department of Psychiatry, Columbia University Irving Medical Center

Address correspondence to: kaylee.bodner@duke.edu

Disclosures: Please see disclosures at the end

**Introduction**

Worldwide, some 40 million adults have Alzheimer's disease (AD) and several hundred million may be at elevated risk for AD by virtue of genetic risk, cerebrovascular disease, comorbid illnesses such as diabetes or depression, mild cognitive impairment (MCI) and/or silent buildup of cortical AD pathology (1). Despite extensive research, there are no pharmacological treatments with more than minimal efficacy for mild AD, and prevention strategies are not established.

The COVID-19 pandemic has transformed mobile health applications and telemedicine from nice to have tools into essential healthcare infrastructure (2-4). This need is particularly great for the elderly who, due to their greater risk for infection, may avoid medical facilities or be required to self-isolate. These are also the very groups at highest risk for cognitive decline. For example, during the COVID-19 pandemic artificially intelligent conversational agents were employed by hospitals and government agencies (such as the CDC) to field queries from patients about symptoms and treatments (3). Digital health tools also proved invaluable to provide neuropsychiatric and psychological self-help to people isolated at home or in retirement centers and nursing homes (2-4).

**Computerized Cognitive Training**

Computerized cognitive training (CCT) is one such application of digital health in which individuals can access gamified, engaging, cognitive exercises from their own computers or mobile devices anytime anywhere (5-15). These exercises can be targeted to improve overall cognition or specific domains (such as learning and memory, attention, speed, executive functioning), as well as daily living skills such as financial knowledge or driving performance (5-15). They can potentially be adjusted based on response via self-administered cognitive tests, and adherence supervised remotely, as needed, by a physician or psychologist. There are over two-dozen CCT programs on the market and consumer demand is large – tens of millions of people in dozens of countries have accessed CCT to date (5).

While popular, scientific opinion regarding CCT has at times been divided. In 2014, two groups with expertise in the field expressed conflicting opinions on the effectiveness of CCT based on the same evidence (6, 7). While one group claimed there was "little evidence that playing brain games improved underlying broad cognitive abilities", the other retorted that "a substantial and growing body of evidence shows that certain cognitive training regimens can significantly improve cognitive function, including in ways that generalize to everyday life." Subsequently, Simons *et al.* criticized many of the industry-sponsored CCT studies on various methodological limitations (8). In a rebuttal, Harvey et al. noted that the critics may have prematurely come to a wrong conclusion – they cited factors such as incorrect definitions of CCT and described supportive evidence from meta-analyses and large RCTs (8). Likewise, while an initial online study by Owen et al (9) did not find any benefits of CCT in younger adults, a subsequent study of 2912 older adults by the same group reported that CCT had benefits on both cognition and daily activities (10).

**Clinical Trials of CCT in Aging and MCI**

The scientific rationale to develop CCT for treating MCI is sound – a large body of observational research suggest that being cognitively active may reduce the risk for dementia and experimental studies show that aging brains retain a capacity for neuroplasticity (reviewed in 5, 11-14).

There is now evidence from relatively large well-controlled trials to support both efficacy and safety. For example, the NIH-funded, Advanced Cognitive Training for Independent and Vital Elderly (ACTIVE) Trial of 2,832 older people, assigned people to 3 forms of training – memory, reasoning and speed – versus a control.  The memory group showed no benefits. But five years after initial training, the reasoning group self-reported fewer daily-living problems, whereas speed-of-processing training resulted in fewer at-fault automobile accidents and a smaller decline in health-related quality of life (12).  Further, at 10-year follow up, those on the computerized speed training arm had a 29% reduction in incident dementia rates (13). A meta-analyses of 52 studies comprising 4,885 cognitively healthy older adults, noted small to moderate beneficial effect sizes for CCT in comparison to control groups in the domains of verbal memory, nonverbal memory, working memory, processing speed, and visuospatial skills (14). This study also found that group-based training was more efficacious than home-based training - suggesting that future home based CCT may need to be augmented with greater remote supervision and interactions via social media (14).   A meta-analysis of 18 studies of CCT for MCI (N=690) found small to moderate improvements in global cognition, memory and working memory (11).  The largest effect size was on working memory.  Whether these improvements result in long-term transfer to clinically meaningful benefits and lowered rates of progression to dementia is not known and require further study. (11).

There is also evidence that the effectiveness of CCT in subjects at risk for AD could be improved by supplementing cognitive training with other tools such as physical exercise, diet, vascular risk reduction, neuromodulation or pharmacotherapy. For example, Lenze et al reported that the addition of a serotonin modulator/stimulator drug, vortioxetine, could improve the efficacy of CCT in MCI (17).  Two studies that examined the effects of combining physical and cognitive training in MCI reported mixed results (15, 16).  Singh et al, using a 2x 2 design, found that CCT improved memory in MCI at 6 months but did not augment the effects of exercise (15).  In contrast, the 40-week population study by Shimada et al (16) of 945 MCI subjects reported that combined CCT and physical exercise improved memory and nonmemory domains, and reduced medial temporal lobe atrophy in amnestic MCI (16).  Lastly, the two-year FINGER randomized controlled trial of 1260 older adults showed that a multi-domain lifestyle intervention, comprising CCT as one of the components, slowed cognitive decline (18).

**Software as a Medical Device**

The International Medical Device Regulators Forum (IMDRF) for software as a medical device (SaMD) proposed consensus guidelines for what constitutes a mobile medical app (i.e. digital therapeutic) versus a wellness app (19). These guidelines stated that a software app intended to treat or prevent a serious disease would have to conduct well-controlled clinical trials to prove efficacy and seek pre-marketing authorization (PMA) from a regulatory agency. Recently, prescription digital therapeutics have been cleared by the US Food and Drug Administration (FDA) for use in substance abuse and sleep disorders, and apps for other diseases are in development (20). Per these guidelines, CCT that is marketed for treating MCI or preventing AD would be viewed as a medical device and subject to pre-marketing regulatory oversight. CCT intended for use as a general wellness tool to improve mental speed would likely not be subject to such oversight.

**Advancing CCT as a Digital Brain Therapeutic**

We believe the most efficient regulatory path for CCT is to seek a marketing indication for the symptomatic treatment of MCI or very mild AD dementia. Such a path would be supported by the large public health threat posed by AD, the high failure rate of investigational drug trials and the urgent need for scalable, low risk, cost-effective, home-based preventive treatments. The small to moderate effect sizes seen in MCI CCT trials to date are likely to be similar to those expected in ongoing anti-amyloid or anti-tau trials. Further, the safety of CCT is superior to most biologics/drugs being studied for MCI and the risk is minimal.

While regulatory approaches for devices often differ from that of drugs, recent FDA draft guidelines for acceptable outcomes in early AD trials of investigational drugs (21) provide a roadmap for CCT. The FDA guidance categorizes early AD into three stages – Stage 1 (pathological changes but no clinical deficits), Stage 2 (mild cognitive deficits but no measurable functional deficits), and Stage 3 (measurable cognitive and functional deficits). Stages 2 and 3 are analogous to early MCI and late-MCI.

The above guidance suggests that in Stage 1 one or more biomarkers could serve as a primary basis for accelerated approval with the requirement for a post-approval confirmatory clinical study. In Stage 2, one or more neuropsychological tests (either effect on multiple tests or a large effect on a single test) could serve as the basis for approval. In Stage 3, a single integrated scale that measures both daily function and cognitive effects (e.g. Clinical Dementia Rating Scale) could serve as evidence of efficacy.

Although the overall literature shows CCT to be safe with a high likelihood of cognitive efficacy in aging and MCI, the lack of positive, regulatory quality trials is the major limitation. The bar for clearance of a software device is often lower than that of a drug; hence it is likely regulatory agencies may view the existing studies of CCT in aging and MCI (such as those cited in 5-17) as supportive and may require only a single, methodologically rigorous, relatively short (e.g. 24-week) trial to gain such an indication. Given the lack of a predicate or product code, CCT for MCI would likely be viewed by the FDA as a Class III device (22); however, we believe that a de-novo application to request re-classification of CCT as a lower risk Class II device could be successful. CCT manufacturers should seek advice from regulatory agencies before and during this process as is done with drugs (19). Alternatively, CCT manufacturers seeking an MCI indication could also utilize the FDA's digital software pre-certification (Pre-Cert) program. Such a path would increase trust among clinicians, consumers and payers.

To date, however, CCT products aimed at the elderly have chosen the less risky path of going directly to consumers as a "wellness" product – suggesting the need for incentives such as market exclusivity or preferential formulary access. Alternatively a regulatory quality trial could also be conducted in the public interest through a public-private partnership involving one or more CCT companies or via a government grant. For example, our group is currently conducting an 18-month randomized trial of CCT versus active control in carefully selected MCI patients with clinically meaningful cognitive (ADAS-Cog), functional (FAQ, UPSA), neuronal loss (hippocampal volume) and disease modifying (progression to dementia) outcomes (23).

Given the millions of elderly already doing CCT at home, it would also be insightful to analyze existing large registries to examine real world outcomes consistent with the FDA's total product lifecycle approach (22). Three areas of real world health analytics (RWHA) would be relevant for CCT – 1) patient reported outcomes such as daily activities; 2) user experience analytics such as engagement and compliance; 3) product performance (reliability, privacy and cybersecurity). Updates on real world performance could be provided quarterly to public and regulators.

For clinicians and the general public to be willing to use CCT as a primary treatment modality, convincing evidence in well-controlled trials with a clear focus on MCI, for example, would be helpful. Future research to clarify the role of augmenting agents, such as off-label medications (e.g. vortioxetine), cholinesterase inhibitors, physical exercise and other non-pharmacologic interventions, for CCT to achieve maximum efficacy as a cognitive enhancing strategy would also be useful. Future studies could also examine it's utility in combination with anti-amyloid or anti-tau agents.

The COVID-19 pandemic has illustrated the demand for digital tools across the entire spectrum of healthcare. Reimbursement and regulatory burdens faced by digital tools before the pandemic have begun to diminish. Post-pandemic, companies that integrate digital therapeutics with other modalities (e.g. digital diagnostics, digital pharmacy, live consults via tele-medicine) will best provide a seamless experience for consumers. A patient-centered, real world health data sharing platform that can collect and aggregate siloed data sources across multiple health systems has recently been demonstrated (24). These lessons are highly relevant to optimize CCT as a clinical tool in MCI.

In summary, we believe that it is an important time for the field to advance CCT from a wellness product to a well-integrated, digital brain therapeutic via an appropriate regulatory pathway. If efficacy is established in MCI, then CCT could be combined with specific self-rated cognitive and functional assessment scales as well as other clinical care options to help millions of elderly both during pandemics and in normal times.

**Acknowledgments**:

The authors are funded by an NIA grant which is evaluating the efficacy of CCT in MCI. KB receives salary support from an NIA grant. TG has received research grants from the NIA and receives royalties for the use of the Brief Assessment of Cognition in Schizophrenia (BACS) in clinical trials. DD has received research grants from the NIH and honoraria as a scientific adviser to Acadia, Genentech, BXCel, Grifols, and Corium. PMD has received grants from the NIH, DOD, ONR, ADDF, Cure Alzheimer's Fund, and the Karen L. Wrenn Trust. PMD has received advisory or board fees from several health and technology companies; PMD owns shares in several companies and is a co-inventor on patents whose products are not discussed here.

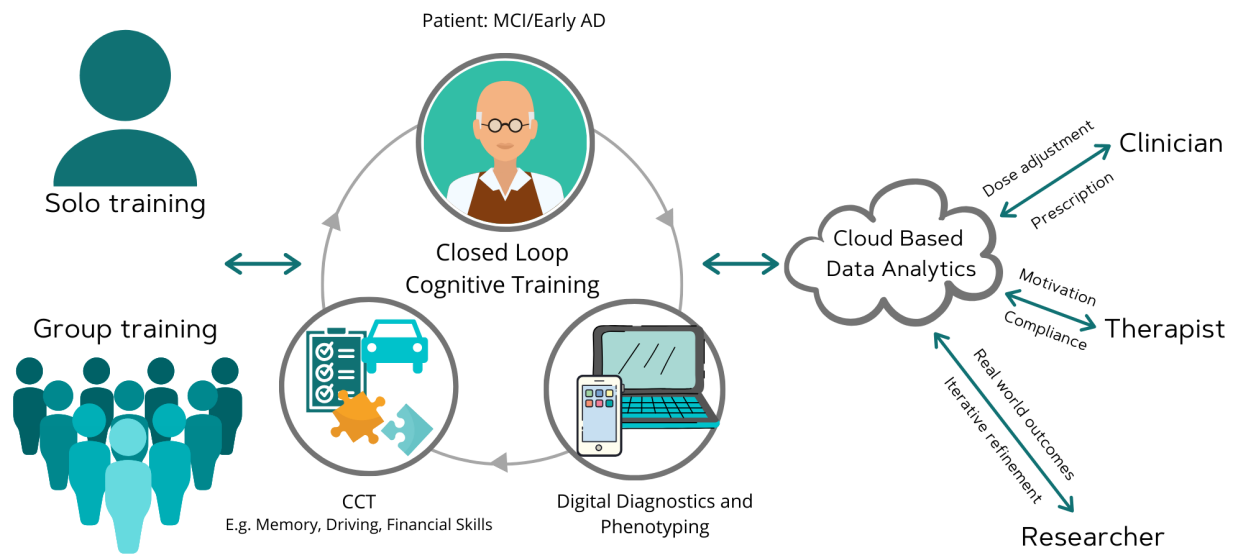

**References:**

1. Dementia statistics. Alzheimer's Disease International. https://www.alz.co.uk/research/statistics. Accessed April 28, 2020.
2. Gordon J, Doraiswamy PM. High anxiety calls for innovation in digital mental health. World Economic Forum, 2020. https://www.weforum.org/agenda/2020/04/high-anxiety-calls-for-innovation-in-digital-mental-health-6b7b4e7044
3. Torous J, Myrick KJ, Rauseo-Ricupero N, Firth J. Digital Mental Health and COVID-19: Using Technology Today to Accelerate the Curve on Access and Quality Tomorrow. *JMIR Mental Health*. 2020;7(3). doi:10.2196/18848.
4. Webster P. Virtual health care in the era of COVID-19. *The Lancet*. 2020;395(10231):1180-1181. doi:10.1016/s0140-6736(20)30818-7.
5. Sternberg DA, Ballard K, Hardy JL, Katz B, Doraiswamy PM, Scanlon M. The largest human cognitive performance dataset reveals insights into the effects of lifestyle factors and aging. *Frontiers in Human Neuroscience*. 2013;7. doi:10.3389/fnhum.2013.00292.
6. A Consensus on the Brain Training Industry from the Scientific Community. Max Planck Institute for Human Development and Stanford Center on Longevity. 2014.

http://longevity.stanford.edu/a-consensus-on-the-brain-training-industry-from-the-scientific-community-2/

7. Harvey PD, Mcgurk SR, Mahncke H, Wykes T. Controversies in Computerized Cognitive Training. *Biological Psychiatry: Cognitive Neuroscience and Neuroimaging*. 2018;3(11):907-915. doi:10.1016/j.bpsc.2018.06.008.
8. Simons DJ, Boot WR, Charness N, Gathercole SE, Chabris CF, Hambrick DZ, Stine-Morrow EAL. Do 'brain training' programs work? *Psychological Science in the Public Interest.* 2016:17(3):103-186.
9. Owen A, Hampshire A, Grahn J *et al.* Putting brain training to the test. *Nature* **465,** 775–778 (2010). https://doi.org/10.1038/nature09042
10. Corbett A, Owen A, Hampshire A, et al. The Effect of an Online Cognitive Training Package in Healthy Older Adults: An Online Randomized Controlled Trial. *Journal of the American Medical Directors Association*. 2015;16(11):990-997. doi:10.1016/j.jamda.2015.06.014.
11. Zhang H, Huntley J, Bhome R, et al. Effect of computerised cognitive training on *cognitive outcomes in mild cognitive impairment: a systematic review and meta-analysis. BMJ* Open. 2019; 9(8): e027062.
12. Tennstedt SL, Unverzagt FW. The ACTIVE study: study overview and major findings. *J Aging Health*. 2013;25(8 Suppl):3S–20S. doi:10.1177/0898264313518133
13. Rebok GW, Ball K, Guey LT, et al. Ten-year effects of the advanced cognitive training for independent and vital elderly cognitive training trial on cognition and everyday functioning in older adults. *J Am Geriatr Soc*. 2014;62(1):16–24. doi:10.1111/jgs.12607
14. Lampit A, Hallock H, Valenzuela M. Computerized Cognitive Training in Cognitively Healthy Older Adults: A Systematic Review and Meta-Analysis of Effect Modifiers. *PLoS Medicine*. 2014;11(11). doi:10.1371/journal.pmed.1001756.
15. Singh MAF, Gates N, Saigal N, et al. The Study of Mental and Resistance Training (SMART) Study—Resistance Training and/or Cognitive Training in Mild Cognitive Impairment: A Randomized, Double-Blind, Double-Sham Controlled Trial. *Journal of the American Medical Directors Association*. 2014;15(12):873-880. doi:10.1016/j.jamda.2014.09.010
16. Shimada H, Makizako H, Doi T, et al. Effects of Combined Physical and Cognitive Exercises on Cognition and Mobility in Patients With Mild Cognitive Impairment: A Randomized Clinical Trial. *Journal of the American Medical Directors Association*. 2018;19(7):584-591. doi:10.1016/j.jamda.2017.09.019.
17. Lenze EJ, Stevens A, Waring JD, et al. Augmenting Computerized Cognitive Training With Vortioxetine for Age-Related Cognitive Decline: A Randomized Controlled Trial. *American Journal of Psychiatry*. 2020. doi:10.1176/appi.ajp.2019.19050561.
18. Ngandu T, Lehtisalo J, Solomon A, et al. A 2 year multidomain intervention of diet, exercise, cognitive training, and vascular risk monitoring versus control to prevent cognitive decline in at-risk elderly people (FINGER): a randomised controlled trial. *The Lancet*. 2015;385(9984):2255-2263. doi:10.1016/s0140-6736(15)60461-5.
19. Center for Devices and Radiological Health. Software as a Medical Device (SaMD). U.S. Food and Drug Administration. https://www.fda.gov/medical-devices/digital-health/software-medical-device-samd. Accessed April 6, 2020.
20. Makin S. The emerging world of digital therapeutics. Nature News. 2019. https://www.nature.com/articles/d41586-019-02873-1.

21. Early Alzheimer's Disease: Developing Drugs for Treatment Guidance for Industry. *US Food and Drug Administration.* 2018:1-7. https://www.fda.gov/media/110903/download.
22. Developing Software Precertification Program. *US Food and Drug Administration*. 2018;2:1-45. https://www.fda.gov/media/113802/download.
23. D'Antonio J, Simon-Pearson L, Goldberg T, et al. Cognitive training and neuroplasticity in MCI: protocol for a 2-site blinded randomized controlled treatment trial. BMJ Open. 2019;9:e028536. doi:10.1136/
24. Dhruva SS, Ross JS, Akar JG, et al. Aggregating Multiple Real-World Data Sources using a Patient-Centered Health Data Sharing Platform: an 8-week Cohort Study among Patients Undergoing Bariatric Surgery or Catheter Ablation of Atrial Fibrillation. February 2019. doi:10.1101/19010348.